\pdfoutput=1
\documentclass[12pt,a4paper]{article}
\usepackage{times}
\usepackage{graphics}
\usepackage{graphicx}
\usepackage{float}
\restylefloat{figure}

\newcommand{\si}{\mbox{\boldmath $\sigma$}}

\begin{document}
%\large
\date{ }

\begin{center}
{\Large Neutron experiments to search for new spin-dependent interactions}

\vskip 0.7cm

Yu. N. Pokotilovski\footnote{e-mail: pokot@nf.jinr.ru; tel:
7-49621-62790; fax: 7-49621-65429}

\vskip 0.7cm
            Joint Institute for Nuclear Research\\
              141980 Dubna, Moscow region, Russia\\
\vskip 0.7cm

{\bf Abstract\\}

\begin{minipage}{130mm}

\vskip 0.7cm
 The consideration is presented of possible neutron experiments to search for
new short-range spin-dependent forces.
 The spin-dependent nucleon-nucleon interaction between neutron and nuclei
may cause different effects: phase shift of a neutron wave in neutron interferometers
of different kind, in particular of the Lloyd mirror configuration, neutron spin rotation
in the pseudo-magnetic field, and transverse deflection of polarized neutron beam by a slab of substance.
 Estimates of sensitivity of these experiments are performed.
\end{minipage}
\end{center}
\vskip 0.3cm


\vskip 0.2cm


\vskip 0.6cm

{\bf 1. Introduction}

 A number of proposals were published for the existence of new
interactions coupling mass to particle spin \cite {Lei,Moh,Hill,Fay,Dob}.
 On the other hand, there are theoretical indications that there may exist
light, scalar or pseudoscalar, weakly interacting bosons.
 Generally the masses and the coupling of these particles to nucleons,
leptons, and photons are not predicted by the proposed models.
 The most attractive solution of the strong CP problem is the existence of a
light pseudoscalar boson - the axion \cite{ax}.
 Current algebra techniques is used to relate the masses and coupling
constants of the axion and neutral pion:
$m_{a}=(f_{\pi}m_{\pi}/f_{a})\sqrt{z}/(z+1)$, where $z=m_{u}/m_{d}=0.56$,
$f_{\pi}\approx 93$\,MeV, $m_{\pi}=135$\,MeV, so that
$m_{a}\approx (0.6\times 10^{10}$\,GeV/$f_{a})$\,meV.
 Here $f_{a}$ is the scale of Peccei-Quinn symmetry breaking.

 The axion coupling to fermions has general view $g_{aff}=C_{f}m_{f}/f_{a}$,
where $C_{f}$ is the model dependent factor.
 The axion may have a priori mass in a very large range, namely
($10^{-12}<m_{a}<10^{6}$) eV.
 The main part of this mass range from both -- low and high mass boundaries --
was excluded in result of numerous experiments and constraints from
astrophysical considerations \cite{PDG}.
 Astrophysical bounds are based on some assumptions concerning the axion and
photon fluxes produced in stellar plasma.

 More recent constraints limit the axion mass to
($10^{-5} <m_{a}< 10^{-3}$) eV with respectively very small coupling
constants to quarks and photon \cite{PDG}.

 Axion is one of the best candidates for the cold dark matter of the Universe
\cite{dark}.
 Some recent reviews are \cite{PDG,rev}.

 Several laboratory searches provided constraints on the product of the scalar
and pseudoscalar couplings at macroscopic distances $\lambda >1$\, mm.
 The examples of the most recent experiments are \cite{Heck,Ni,Ritt,Hamm}.

 These experiments may be divided in two groups: measurement of the force
between a mass and a sensor, or measurement of the frequency shift
resulting from interaction between mass and a sensor (Fig. 1).

 In the experiments, where neutron is used as a sensor, in addition to these two possibilities
the phase shift of the neutron wave function due to spin-dependent interaction may be measured.

 On the other hand there exist independent constraints on the scalar and
pseudoscalar coupling constants.
 The first one follows from the experiments of Galileo-, E\"otv\"os- or
Cavendish-type performed with macro-bodies, in measurements of the Casimir
force between closely placed macro-bodies, or from high precision particle
scattering.
 For the interaction range of our interest: 1 cm  $>\lambda >10^{-4}$ cm
the most sensitive are the  measurements of Seattle and Stanford groups
\cite{Se04,Se07,Seconstr,St03,St05,St08}.
 They have searched a deviation from the Newton gravitation law
%1
\begin{equation}
U(r)=-\frac{GMm}{r}(1+\alpha e^{-r/\lambda})
\end{equation}
where $G$ -- the Newtonian gravitational constant, $M$ and $m$ are the masses
of gravitating bodies.

 For the monopole-monopole interaction due to exchange of the pseudo-scalar
boson
%2
\begin{equation}
V_{mon-mon}(r)=\frac{g_{s}^{2}}{4\pi}\frac{\hbar c}{r}e^{-r/\lambda}
\end{equation}
the scalar coupling constant $g_{s}$ may be inferred from the limits on
$\alpha$:
%3
\begin{equation}
g_{s}^{2}=\frac{4\pi Gm_{n}^{2}\alpha}{\hbar c}\approx 10^{-37}\alpha,
\end{equation}
where $m_{n}$ is nucleon mass.

 It follows from the experiments \cite{Se04,Se07,St03,St05,St08} that
$g_{s}^{2}$ is limited by the value $10^{-40}-10^{-38}$ in the interaction
range 1 cm $>\lambda >10^{-4}$ cm (see also \cite{Seconstr}).

 The pseudoscalar coupling constant is restricted to $g_{p}<10^{-9}$ from
astrophysical considerations \cite{Raf}.

 Sensitivity of these experiments falls with decreasing the interaction range below $\sim$1 cm.

 The experimental limit on the monopole-dipole interaction in the
$\lambda$--range $(10^{-4}-1)$ cm was established in the Stern-Gerlach type
experiment in which ultracold neutrons transmitted through a slit
between a horizontal mirror and absorber \cite{grav}.
 Their limit for the value $g_{s}g_{p}$ is $\sim 10^{-15}$ at the
$\lambda=10^{-2}$ cm, what corresponds to the value of the monopole-dipole
potential at the surface of the mirror $\sim 10^{-3}$ neV, which is
equivalent to the magnetic field $\sim$0.2 G in the neutron magnetic
moment interaction ${\bf\mu H}$ with magnetic field.

 Sensitivity estimates for a future ultracold neutron Stern-Gerlach type
experiment were presented, which promise orders of magnitude improvements
in limiting the monopole-dipole interaction \cite{grav1}.

 Constraints on new spin-dependent interactions were also obtained from
existing data on ultracold neutron depolarization in traps \cite{Ser,myUCN}.

 There was also a proposal of the ultracold neutron magnetic resonance
frequency shift experiment for obtaining these constraints with better
precision \cite{Zimm}.
 The realization of this idea is described in \cite{Ser-Zimm}.

 Analysis of the longitudinal spin relaxation of polarized $^{3}He$ to constrain axion-like interactions
was proposed in \cite{myHe} (see also similar constraints obtained in \cite{Fu} and corrections in \cite{Pet}).

 It is seen that the constraints obtained and expected from further laboratory searches are weak
compared to the limit on the product $g_{s}g_{p}<10^{-28}$ inferred from the above mentioned
separate constraints on $g_{s}$ and $g_{p}$.
 Although laboratory experiments may not lead to the strongest bounds numerically, measurements
made in terrestial laboratories produce the most reliable results.
 On the other hand the direct experimental constraints on the
monopole-dipole interaction may be useful for limiting more general class of
low-mass bosons irrespective to any particular theoretical model.
 Further the constraint on product $g_{s}g_{p}$ may be used for the limits on the
coupling constant of this more general interaction.

 Axions mediate a P- and T-reversal violating monopole-dipole
interaction potential between spin and matter (polarized and
unpolarized nucleons) \cite{Mood}:
%4
\begin{equation}
V_{mon-dip}({\bf r})=g_{s}g_{p}\frac{\hbar^{2}\si\cdot {\bf n}}
{8\pi m_{n}}\Bigl(\frac{1}{\lambda r}+\frac{1}{r^{2}}\Bigr)e^{-r/\lambda},
\end{equation}
where $g_{s}$ and $g_{p}$ are dimensionless coupling constants of
the scalar and pseudoscalar vertices (unpolarized and polarized
particles), $m_{n}$ the nucleon mass at the polarized vertex, the
nucleon spin ${\bf s}=\hbar\si/2$, $r$ is the distance
between the nucleons, $\lambda=\hbar/(m_{a}c)$ is the range of the
force, $m_{a}$ - the axion mass, and $\bf n={\bf r}$/r is the
unitary vector.

 The potential between the layer of substance and the nucleon separated by the
distance $x$ from the surface is:
%5
\begin{equation}
V_{mon-dip}(x)=\pm g_{s}g_{p}\frac{\hbar^{2}N\lambda}{4m_{n}}
(e^{-|x|/\lambda}-e^{-|x+d|/\lambda})=V_{0}e^{-|x|/\lambda} \qquad (d\gg\lambda),
\end{equation}
where $N$ is the nucleon density in the layer, $d$ is the layer's thickness.
 The "+" and "-" depends on the nucleon spin projection on x-axis (the surface
normal).

 We consider different neutron experiments to search for or to constrain this interaction.
%===================================================================
\vspace{5mm}

 {\bf 2. Phase shift in the neutron interferometer.}

 Slow neutrons with polarization normal to the surface of unpolarized matter
(Fig. 2) experience shift of the phase of the wave function:
%6
\begin{equation}
\beta=\frac{2l V(x)}{\hbar v}= g_{s}g_{p}\frac{N\lambda\lambda_{n}l}
{4\pi}e^{-x/\lambda}\longrightarrow g_{s}g_{p}\frac{N\lambda\lambda_{n}l}
{4\pi},
\end{equation}
where $\lambda_{n}$ is the neutron wavelength, $l$ is the length of the
inserted sample, $x$ is the distance of the beam from the surface.
 We assume here that $x$ - distance of the beam from the surface is less than
the interaction range $\lambda$ and both $x,\lambda\ll d$.

 The sensitivity to the product of the coupling constants follows from the
sensitivity to the measured phase shift:
%7
\begin{equation}
g_{s}g_{p}=\frac{4\pi\beta}{N\lambda\lambda_{n}l}.
\end{equation}

 At the very low energy interferometer with $\lambda_{n}=30$ \AA\,
\cite{interf}, $\lambda$=1 mm, $l=$1 cm, $\beta=10^{-3}$, $N=5\times 10^{24}$,
we have $g_{s}g_{p}=10^{-21}$.
%==================================================================
\vspace{5mm}

 {\bf 3. Neutron spin rotation.}

The neutron spin Larmor precession frequency is
%8
\begin{equation}
\omega_{L}=\frac{2V(x)}{\hbar}=g_{s}g_{p}\frac{\hbar N\lambda}{2m_{n}}
e^{-|x|/\lambda}.
\end{equation}
 Angle of spin rotation
%9
\begin{equation}
\varphi=\omega_{L}t=\omega_{L}\frac{l}{v}=
g_{s}g_{p}\frac{N\lambda\lambda_{n}l}{4\pi}e^{-|x|/\lambda},
\end{equation}
the same as in the interferometer, but the sensitivity to the $g_{s}g_{p}$
may be higher.

 It follows that constraint on $g_{s}g_{p}$
%10
\begin{equation}
g_{s}g_{p}=\frac{4\pi\varphi}{N\lambda\lambda_{n}l}e^{|x|/\lambda}.
\end{equation}
 At $\lambda_{n}=10\, \AA$, $N\sim 5\times 10^{24}$, $l= 1\, m$, $\lambda$=1 mm,
and measurable spin rotation angle $\varphi=10^{-5}$\, the limit may be obtained
$g_{s}g_{p}=3\times 10^{-23}e^{|x|/\lambda}$.
%======================================================================
\vspace{5mm}

{\bf 4. Deviation of polarized neutron beam.}

 The gradient of the potential of Eq. (5) is
%11
\begin{equation}
\nabla V_{mon-dip}(x)=g_{s}g_{p}\frac{\hbar^{2}N}{4m_{n}}e^{-|x|/\lambda}
\end{equation}
and is directed along the nucleon spin direction ($\lambda\ll d$).

 The acceleration exerted by the medium on the polarized neutron is
%12
\begin{equation}
a=g_{s}g_{p}\frac{\hbar^{2}N}{4m_{n}^{2}}e^{-|x|/\lambda}.
\end{equation}

 For an estimate of the sensitivity of such an experiment we consider the
experiment \cite{Bau} on the search for the neutron electric charge.
 In this experiment the beam of cold neutrons with the wavelength
($12<\lambda< 30$) \AA\, passed through a strong electric field $\sim$ 60 kV/cm, over a length of 9 m, and
possible lateral deflection of the beam was looked with a multi-slit neutron optical system when polarity of the
electric field was changed alternatively.

 In the search for the spin-dependent interaction the expected lateral
deflection angle of the neutron beam with velocity $v$ polarized normal to the
surface of the slab of substance with the length $l$
%13
\begin{equation}
\theta=\frac{v_{\perp}}{v_{\parallel}}=\frac{at}{v_{\parallel}}=
\frac{al}{v_{\parallel}^{2}}=
g_{s}g_{p}\frac{\hbar^{2}Nl}{4m_{n}^{2}v_{\parallel}^{2}},
% g_{s}g_{p}\frac{\lambda_{n}^{2}Nl}{(4\pi)^{2}},
\end{equation}
when $x\ll\lambda$.
 Similarly, in the neutron charge experiment
%14
\begin{equation}
\theta=\frac{v_{\perp}}{v_{\parallel}}=\frac{at}{v_{\parallel}}=
\frac{al}{v_{\parallel}^{2}}=
\frac{Q_{n}El}{m_{n}v_{\parallel}^{2}}.
\end{equation}

 From the constraints for $Q_{n}$ the sensitivity of a possible experiment for
the constraints on $g_{s}g_{p}$ is:
%15
\begin{equation}
g_{s}g_{p}=\frac{4m_{n}Q_{n}E}{\hbar^{2}N}
\end{equation}
 At $Q_{n}=10^{-21}e$  \cite{Bau} and $N=5\times 10^{24}$ cm$^{-3}$ we get
$g_{s}g_{p}=10^{-22}$.

 Serious requirement in this experiment is perfect magnetic shielding:
the lateral deflection of the neutron beam caused by the pseudo-magnetic field at the value
$g_{s}g_{p}=10^{-22}$  may be caused by the gradient of magnetic field of $\sim 10^{-5}$ G/cm directed normal to
the slab surface.

%======================================================================
\vspace{5mm}

{\bf 5. Lloyd mirror neutron interferometer.}

 The Lloyd mirror neutron interferometer, which to the author's knowledge never was tested before,
can provide high sensitivity to spin-dependent forces.
 Fig. 3 shows the geometry.

 The neutron wave vector $k^{'}$ in the potential $V$ is
%16
\begin{equation}
k^{'2}=k^{2}-\frac{2mV}{\hbar^{2}}; \qquad k^{'}=k-\frac{mV}{k\hbar^{2}},
\end{equation}
where $k$ is the neutron wave vector in the absence of potential.

 The phase shift due to interaction potential of the neutron with the mirror depending on distance
from the mirror is obtained by integration along trajectories $\Delta\varphi=\oint k^{'}
ds=\varphi_{I}-\varphi_{II}$, where $\varphi_{I}$ and $\varphi_{2}$ are phases obtained due to potential of Eq.
(5) along trajectories I and II respectively:
%17
\begin{eqnarray}
\varphi_{I} & = & k\sqrt{L^{2}+(b-a)^{2}}+\frac{mV_{0}}{k\hbar^{2}L}\sqrt{L^{2}+(b-a)^{2}}
\int\limits_{0}^{L}exp\bigl[-(a+\frac{b-a}{L}x)/\lambda\bigr]dx= \nonumber\\
& & =\varphi_{geom.I}+ \frac{mV_{0}}{k\hbar^{2}}\sqrt{L^{2}+(b-a)^{2}}\frac{\lambda}{b-a}
(e^{-a/\lambda}-e^{-b/\lambda})=\varphi_{geom.I}+\varphi_{pot.I}
\end{eqnarray}
and

%18
\begin{eqnarray}
\varphi_{II}=k\sqrt{L^{2}+(b+a)^{2}}+\frac{mV_{0}}{k\hbar^{2}L}\sqrt{L^{2}+(b+a)^{2}}
\Bigl[\int\limits_{0}^{l}exp\bigl[-(a-\frac{b+a}{L}x)/\lambda\bigr]dx+ \nonumber\\
\int\limits_{l}^{L}exp\bigl[-(\frac{b+a}{L}x-a)/\lambda\bigr]dx\Bigr]=
\varphi_{geom.II}+\frac{mV_{0}}{k\hbar^{2}}\sqrt{L^{2}+(b+a)^{2}}\frac{\lambda}{b+a}
(2-e^{-a/\lambda}-e^{-b/\lambda})=\nonumber\\
=\varphi_{geom.II}+\varphi_{pot.II}.
\end{eqnarray}
 In Eq. (18) $l=aL/(a+b)$ is $x$-coordinate of the reflection point.

 With good precision $\varphi_{geom.I}-\varphi_{geom.II}=2kab/L$.

 For spin-dependent potential of Eq.(5) signs of potential $V_{0}$ and,
respectively, the phase shift $\varphi_{pot}$ are opposite for two spin orientations in respect to the mirror
surface normal.

 The difference in these phase shifts, measured in the experiment
%19
\begin{equation}
(\varphi_{I}^{+}-\varphi_{II}^{+})-(\varphi_{I}^{-}-\varphi_{II}^{-})=
2(\varphi_{I}-\varphi_{II})=\delta\varphi.
\end{equation}

 For non-perfectly vertical mirror the component of gravity normal to the
surface of the mirror produces the potential $V_{gr}=cmgz$, where $g$ is gravitational acceleration and the
coefficient $c$ depends on the angle $\theta$ between gravity vector and the mirror plane.
 At $\theta=1^{'}$ $c\approx 3\times 10^{-4}$.
 This linear potential leads to additional phase shift
%20
\begin{equation}
\delta\varphi_{gr}=\frac{cgm^{2}}
{2k\hbar^{2}(a+b)}\Bigl[(b+a)^{2}\sqrt{L^{2}+(b-a)^{2}}-
(b^{2}+a^{2})\sqrt{L^{2}+(b+a)^{2}}\Bigr],
\end{equation}
calculated in analogy with Eqs.(17-18).

 Finally we should calculate the phase shift of the neutron wave along the
beam II at the point of reflection.
 Neglecting imaginary part of the potential of the mirror, the amplitude of
reflected wave is $r=e^{-i\Theta}$, with the phase $\Theta=2 \arccos(k_{norm}/k_{b})
\approx \pi-2k_{norm}/k_{b}$, where $k_{norm}$ is normal to the mirror surface component
of the neutron wave vector, and $k_{b}$ is the boundary wave vector of the mirror.

 These two phase shifts do not depend on spin.

 Fig.4 shows the neutron wave phase shift $\delta\varphi$ at different interaction range $\lambda$.
 Lloyd mirror interferometer has parameters: $L$=1 m, $a=50\, \mu m$, neutron
wave length 100 \AA \, (neutron velocity 40 m/s), and $g_{s}g_{p}=10^{-19}$; 1 - gravitational phase shift
$\delta\varphi_{gr}$ at $c=3\times 10^{-4}$; 2 - $\pi-$phase shift of the ray II at reflection.

% Fig.5 demonstrates similar results for geometry with $a=20 \mu m$.

 In principle, similar experiments may be performed to search for interactions of the type of Eq. (1).

 Similarly to Eq. (5) additional Yukawa-type interaction of Eq. (1) yields potential
 between the layer of substance and the nucleon separated by the distance $x$ from the surface:
%21
\begin{equation}
U(x)=\alpha 2\pi NG\lambda^{2}m_{n}^{2}e^{-|x|/\lambda}.
\end{equation}

The ratio of potentials of Eqs. (21) and (5) is
%22
\begin{equation}
\frac{U}{V_{mon-dip}}=\frac{\alpha}{g_{s}g_{p}}\frac{m_{n}^{3}G\lambda}{\hbar^{2}} \approx 3\cdot
10^{-25}\lambda \mbox { (cm) } \frac{\alpha}{g_{s}g_{p}}.
\end{equation}

 At $\lambda=0.01$\, cm we obtain $\frac{U}{V_{mon-dip}}\approx 3\cdot 10^{-27}\frac{\alpha}{g_{s}g_{p}}$.
 At equal experimental sensitivity to potentials we have $\alpha\approx 3\cdot 10^{26}g_{s}g_{p}$.

%=================================================================
\newpage

\begin{figure}
\begin{center}
\resizebox{15cm}{4cm}{\includegraphics[width=\columnwidth]{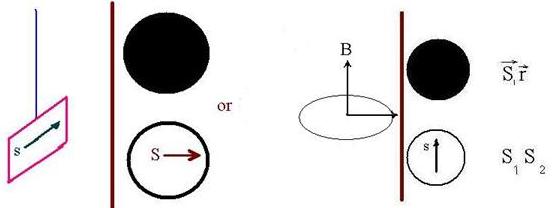}}
\end{center}
\caption{Scheme of macroscopic experiments to search for spin-dependent
forces by: left -- acceleration method, right -- the frequency shift method.}
\end{figure}

\newpage

\begin{figure}
\begin{center}
\resizebox{15cm}{3cm}{\includegraphics[width=\columnwidth]{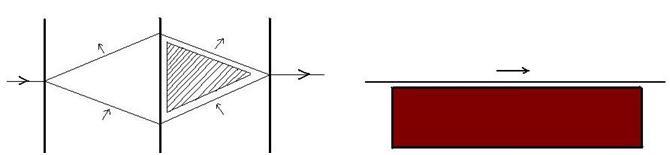}}
\end{center}
\caption{Left -- scheme of the interferometric experiment to search for new spin-dependent forces, right --
neutron spin rotation or beam deflection experiment.}
\end{figure}

\newpage

\begin{figure}
\begin{center}
\resizebox{12cm}{6cm}{\includegraphics[width=\columnwidth]{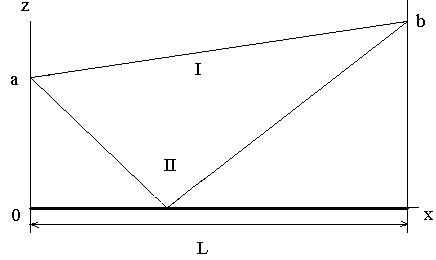}}
\end{center}
\caption{Lloyd neutron interferometer.
 Reflecting plane is vertical so that gravity effect on interference is reduced.
 The height of the slit above the reflecting plane is $a$, $L$ is the distance
from the slit to the detector surface, $b$ is the height of the detector coordinate from the reflecting plane.}
\end{figure}

\newpage

\begin{figure}
\begin{center}
\resizebox{15cm}{13cm}{\includegraphics[width=\columnwidth]{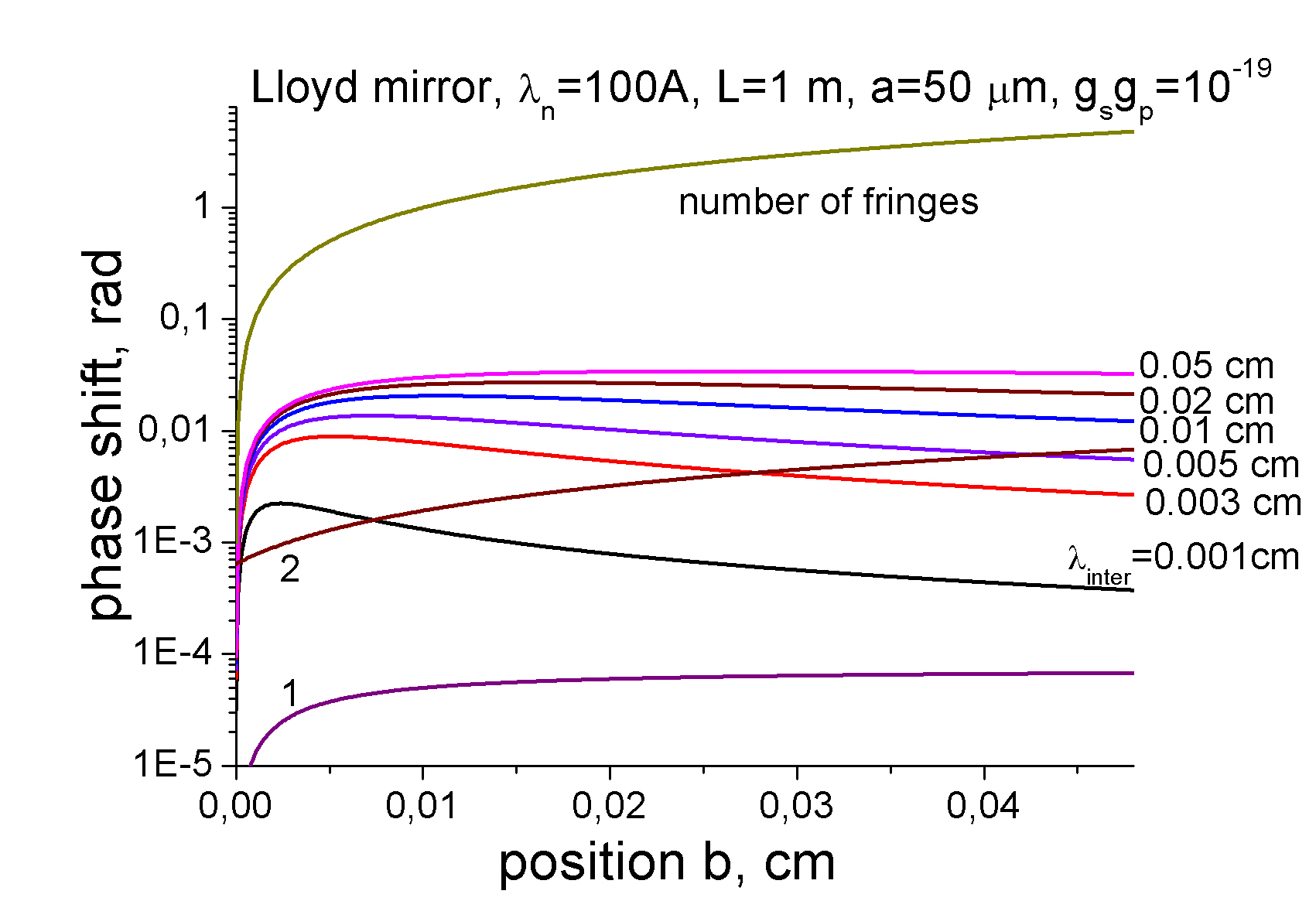}}
\end{center}
\caption{Neutron wave phase shift $\delta\varphi$ at different interaction range $\lambda$.
 Lloyd mirror interferometer has parameters: $L$=1 m, $a=50\,\mu m$, neutron
wave length 100\,\AA \, (neutron velocity 40 m/s), and $g_{s}g_{p}=10^{-19}$; 1 - gravitational phase shift
$\delta\varphi_{gr}$ at $c=3\times 10^{-4}$; 2 - $\pi-$phase shift of the ray II at reflection
($k_{b}=10^{6}$\,cm$^{-1}$).}
\end{figure}

\end{document}